\documentclass{PoS}
\title{Low-mode averaging for  baryon correlation functions
\thanks{Preprint: CERN-PH-TH/2005-172, CPT-2005/P.046} }
\author{Leonardo Giusti \\

        CERN- PH-TH Division,
CH-1211 Geneva, 
Switzerland \\
E-mail: \email{Leonardo.Giusti@cern.ch}}

\author{\speaker{Silvia Necco}\thanks{Supported by TMR, EC-Contract
    No. HPRNCT-2002-00311 (EURIDICE)}\\
        Centre de Physique Th\'eorique{\thanks{UMR 6207 du CNRS et des
        Universit\'es d'Aix-Marseille I,II et du Sud Toulon-Var, affili\'ee \`a
        la FRUMAM}}, 
        Case 907, CNRS Luminy, F-13288
        Marseille Cedex 9, France\\
        E-mail: \email{necco@cpt.univ-mrs.fr}}

\abstract{The low-mode averaging technique is a powerful tool for reducing
large fluctuations in correlation functions due to low-mode
eigenvalues of the Dirac operator. In this work we propose a generalization
to baryons and test our method on two-point correlation functions of
left-handed nucleons, computed with quenched Neuberger fermions on a lattice with
extension $L\;=\;1.5$ fm.
We show that the statistical fluctuations can be reduced
and the baryon signal significantly improved.}

\FullConference{XXIIIrd International Symposium on Lattice Field Theory\\

                 25-30 July 2005\\

                 Trinity College, Dublin, Ireland}

\ShortTitle{Low mode averaging for bayon correlation functions}
\PoS{PoS(LAT2005)132}
\usepackage{epsfig}
\begin{document}

%%%%%%%%%%%%%%%%%%%%%%%%%%%%%%%%%%%%%%%%%%%%%%%%%%%%%%%%%%%%%%%%%%%%%%%%%%%%%%%
\section{Introduction}
It is a known fact that arbitrarily small eigenvalues of the massless Dirac
operator $D$ can occur \cite{Leutwyler:1992yt,Shuryak:1992pi,Damgaard:2001ep,Damgaard:1999tk,Bietenholz:2003mi,Giusti:2003gf}, and that local ``bumps'' of the corresponding wave functions can be reflected in large
fluctuations of the physical observables \cite{Giusti:2003gf}.
In particular, in the quark mass region $m\sim 1/\Sigma V$, where $V$ is the volume and $\Sigma$ is
the bare quark condensate, we expect that few eigenvectors could give a substantial
contribution to the observables; in that case the fluctuations can be reduced in
an efficient way by applying an exact low-mode averaging procedure.
By adopting the techniques developed in \cite{Giusti:2002sm,Giusti:2004yp}, low-mode
averaging has been applied to quenched meson correlation functions with
Neuberger fermions, both in the
$\epsilon$- and $p$-regimes \cite{Giusti:2004yp,DeGrand:2004qw}. \\
This technique can be generalized to a wider class of correlators; in
particular this work is devoted to its application to baryon two-point functions. 
After clarifying the theoretical framework, we present a numerical study 
with quenched Neuberger fermions in the $p$-regime. 
%%%%%%%%%%%%%%%%%%%%%%%%%%%%%%%%%%%%%%%%%%%%%%%%%%%%%%%%%%%%%%%%%%%%%%%%%%%%%%
\section{Low-mode averaging for baryonic two-point functions}
In the following we consider a lattice of volume $V=L^3T$ with lattice spacing
$a$  and periodic boundary conditions in all directions.
We assume that fermions are discretized by using the Neuberger--Dirac operator
$D$ \cite{Neuberger}, which satisfies the Ginsparg--Wilson relation \cite{Ginsparg:1981bj}; this ensures that chiral symmetry is preserved at finite lattice spacing \cite{Luscher:1998pq}. 
The conventions used in this work are the same as in \cite{Giusti:2004yp}, to
which the reader can refer for undefined notations. We adopt the neutron interpolating field
\begin{equation}\label{interp}
N_L(x)=\left[\tilde{u}_{L}^{a T}(x)C\gamma_5\tilde{d}_{L}^b(x)\right]\tilde{d}_{L}^c(x)\epsilon^{abc},
\end{equation}
where $\tilde q=\left(1-\frac{1}{2}\overline{a}D\right)q$ and
$C$ is the charge conjugation matrix. We consider the two-point function 
\begin{equation}\label{nucleocorr}
C_N(t)=\sum_{\vec{x}}\langle{\rm Tr}\left(N_{L}(x) \overline{N}_{L}(0)\gamma_0\right)\rangle,
\end{equation}
where the trace is meant over the nucleon spinor indices.
Following ref. \cite{Giusti:2002sm}, for each gauge
configuration we can extract the first
$n$ low modes of the Dirac operator and express the left--left propagator as
the sum of the \emph{light} and \emph{heavy} parts:
\begin{equation}\label{left_prop}
S_L(x,y)=
P_-S(x,y)P_+= P_- \left[\sum_{i=0}^{n-1}\varphi_i(x)\varphi_i^{\dagger}(y)+S^h(x,y)\right]P_+,
\end{equation}
where
$P_{\pm}=(1\pm\gamma_5)/2$ and
\begin{equation}
\varphi_i(x)=\frac{1}{\sqrt{\overline{\lambda_i}}}\left[
  P_c\psi_i+P_{-c}D P_c\psi_i \right](x).
\end{equation}
Here, $P_c$ ($c=\pm$) is the projector on the chiral sector with no zero modes, and
$\overline{\lambda_i}$, $\psi_i$ are respectively the approximate eigenvalues
and eigenvectors of $A=P_cD_m^{\dagger}D_mP_c$, where $D_m$ is the massive
Neuberger--Dirac operator.
The use of the left-handed propagator avoids the complications due
to the contribution of the zero modes to the correlation function.
The baryon correlation function in eq. (\ref{nucleocorr}) can be split into four contributions:
\begin{equation}
C_N(t)=C_N^{hhh}(t)+C_N^{hhl}(t)+C_N^{hll}(t)+C_N^{lll}(t),
\end{equation}
where $h$ ($l$) means that the heavy (light) part of the propagator is
involved. The ensemble average of each contribution is translational
invariant, and therefore its Monte Carlo variance can be reduced by averaging
over equivalent space-time points.
There are two possible contractions that contribute to the correlation function
of eq. (\ref{nucleocorr}), and in the
 case of degenerate quark masses, the $lll$ part can be written as
\begin{equation}\label{lll}
C_N^{{lll}}(t)=\frac{1}{V}\sum_{x,y}\delta_{t,t_x-t_y}\sum_{i,j=0}^{n-1}
{\rm Tr} \left[P_+\left(\varphi_i^a(x)\varphi_i^{g\;\dagger}(y)\right)^T C\gamma_5 P_-
  \varphi_j^b(x)\varphi_j^{f\;\dagger}(y)C\gamma_5 \right]
\end{equation}
$$
\sum_{k=0}^n {\rm Tr} \left[
  P_-\varphi_k^c(x) \varphi_k^{e\;\dagger}(y)\gamma_0\right]\varepsilon^{abc}\varepsilon^{feg},
$$
where the translational invariance can be fully exploited without any
additional computational cost. 
For the $hll$ contribution, the direct application of low-mode averaging would
require the inversion of the heavy propagator on ${O}(n^2)$ source
vectors. The strategy is then to select in a  
systematic manner only a restricted number of contributions on
which space-time averaging will be applied.
The first step in this direction is to introduce diquark vectors
\begin{equation}
L_{a,\alpha\beta}^{i}(x)  \equiv 
\varepsilon^{abc}\varphi_{i_1,\alpha}^b(x)
\varphi_{i_2,\beta}^c(x),
\end{equation}
where $i=i_1+n\; i_2$, and $\alpha,\beta$ are the explicit spinor
indices. In the space spanned by the $L^i$ we define the scalar product 
\begin{equation}\label{scalar_prod}
(L^i,L^j)=\sum_{x,a,\alpha,\beta,\alpha',\beta'}
L_{a,\alpha\beta}^{\dagger\;i}(x)(P_c)_{\alpha\alpha'}(P_c)_{\beta\beta'}L_{a,\alpha'\beta'}^{j}(x),
\end{equation}
and we define an orthonormal basis $V^i$ by diagonalizing the metric matrix $(L^i,L^j)$.
The basic building block
\begin{equation}\label{bb}
\mathfrak{D}(x,y)=\sum_{k=0}^{n^2-1}L^k(x)L^{k\;\dagger}(y)
\end{equation}
can be rewritten as
\begin{equation}\label{bb_diag}
\mathfrak{D}(x,y)=\sum_{k=0}^{n^2-1}|c_k|^2 V^k(x)V^{k\;\dagger}(y),
\end{equation}
where $|c_0|>|c_1|>...>|c_{n^2-1}|$ are appropriate coefficients.
The $hll$ contribution can be expressed as
\begin{equation}\label{hll}
C_N^{{hll}}(t)=\frac{1}{L^3}\sum_{x,\vec{y}}\Bigg\{
-2\sum_{k=0}^{n^2-1}|c_k|^2{\rm Tr} \left[\left(P_-S^h(x,y)C\gamma_5P_+V^{k\;\dagger}(y)\right)^T C\gamma_5 P_-V^k(x)P_-\gamma_0\right]+
\end{equation}
$$
\sum_{k=0}^{n^2-1}|c_k|^2{\rm Tr}\left[\left({P_-S^h(x,y)P_+{\rm Tr}(C\gamma_5 V^{k\;\dagger}(y)P_+)}\right)^T\gamma_0\right]{\rm Tr}
\left[V^k(x)P_-C\gamma_5 \right]\Bigg\}\delta_{t,t_x-t_y},
$$
and low-mode averaging is applied only on the (supposedly) dominant 
contributions ($k=0,...,n_b-1$), while the remaining terms ($k=n_b,...,n^2-1$)
are computed locally. The other contributions, $hhl$ and $hhh$, are computed
locally as well, under the assumption that their fluctuations are much milder
than in the previous two cases.
%%%%%%%%%%%%%%%%%%%%%%%%%%%%%%%%%%%%%%%%%%%%%%%%%%%%%%%%%%%%%%%%%%%%%%%%%%%
\begin{figure}
\begin{center}
\epsfig{file= 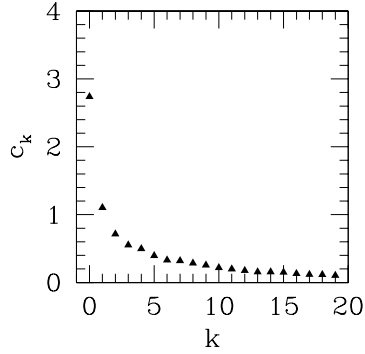,width=5cm}
\caption{The coefficient $c_k$ of eq. (\protect\ref{bb_diag}) as a function of $k$,
  for a given configuration at $am=0.04$. } 
\label{ck}
\end{center}
\vspace{-0.5cm}
\end{figure}
%%%%%%%%%%%%%%%%%%%%%%%%%%%%%%%%%%%%%%%%%%%%%%%%%%%%%%%%%%%%%%%%%%%%%%%%%%%%%%
\begin{figure}
\epsfig{file=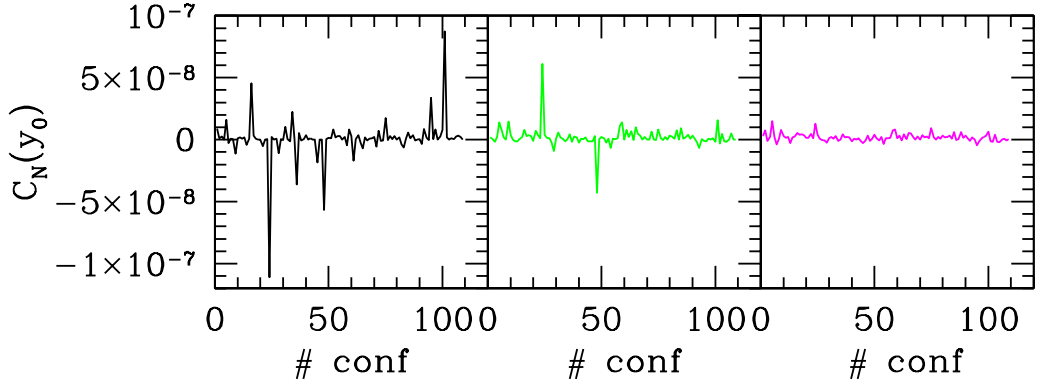,width=14cm}
\caption{Monte Carlo history of the correlation function $C_n(t)$ at
  $t=y_0=9a$, for $am=0.04$.} 
\label{MC_hist}
\vspace{-0.3cm}
\end{figure}
%%%%%%%%%%%%%%%%%%%%%%%%%%%%%%%%%%%%%%%%%%%%%%%%%%%%%%%%%%%%%%%%%%%%%%%%%%%%
\section{Numerical results and discussion}
We tested our method on quenched Neuberger fermions defined on a lattice with
$L/a=12$, $T/a=32$, $\beta=5.8485$, corresponding to $a\simeq 0.12$ fm,  $L\simeq 1.5$ fm.
We used three different quark masses $am=0.04,0.055,0.07$ and collected
$N_{\rm{conf}}=109$ measurements of the baryon correlation function.
We extracted $n=20$ low modes of the Dirac operator, and chose $n_b=20$ for
the low-mode averaging on the $hll$ contribution. Taking into account the symmetries of
the source vectors $V^k$, this corresponds to computing $2n_b$
inversions of the propagator.
In fig. \ref{ck} we show the coefficients $c_k$ of eq. (\ref{bb_diag}) for
$k=0,...,n_b-1$ for a typical configuration at $am=0.04$; they drop off
considerably fast, which justifies the restriction $k=0,...,n_b-1$ for the application of low-mode averaging.
In any case we stress that we performed no systematic study to optimize the choice of the parameter $n_b$.
In fig. \ref{MC_hist} we show the Monte Carlo history of $C_N(t)$ at a given
time slice $t=y_0=9a$, for the lightest mass $am=0.04$. The first figure from
the left corresponds to the local computation, and the presence of large fluctuations can be observed. 
%%%%%%%%%%%%%%%%%%%%%%%%%%%%%%%%%%%%%%%%%%%%%%%%%%%%%%%%%%%%%%%%%%%%%%%%%%%%%%%%%%%%%%%
\begin{table}[h]
\begin{center}
\begin{tabular}{|c|c|c|c|}
\hline
$am$   &  $am_{eff}$ (no lma) & $am_{eff}$ (lma on $lll$) & $am_{eff}$ (lma on $lll$+$hll$) \\
\hline                  
0.04   &   0.69(62)      & 0.76(14)         &   0.78(7) \\
0.055  &   0.89(20)      & 0.81(11)         &   0.85(6) \\
0.07   &   0.94(11)      & 0.87(8)          &   0.90(5) \\
\hline   
\end{tabular}
\caption{Numerical results of low-mode averaging (lma) for the effective mass in units of lattice spacing
  at $t/a=9$,
  for different values of the quark mass $am$.}
\end{center}
\label{masstab}
\vspace{-0.5cm}
\end{table}
\begin{figure}[h]
\epsfig{file=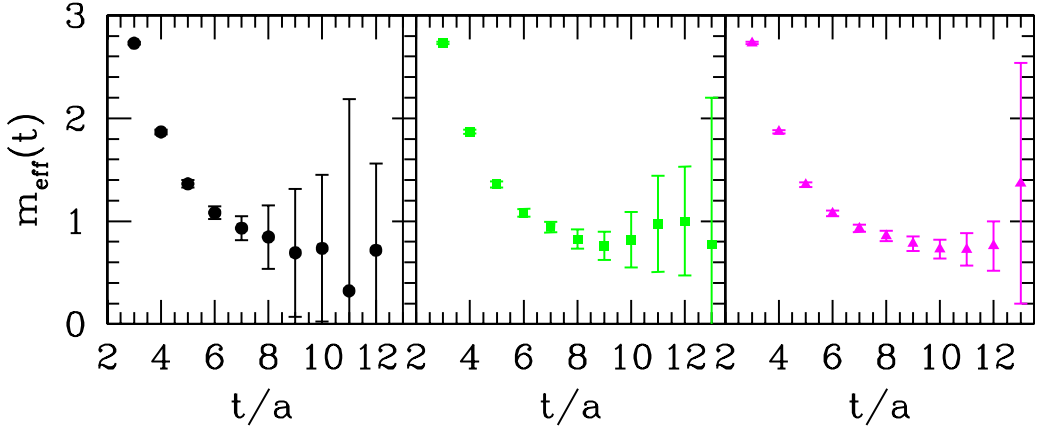,width=14cm}
\caption{Effective mass for $am=0.04$.} 
\label{eff_mass}
\end{figure}
The figure in the centre is obtained
by performing low-mode averaging on the $lll$ part of the correlation
function, as in eq. (\ref{lll}): the fluctuations are reduced, but
nevertheless we still observe residual spikes. Finally, the third curve on
the right corresponds to the computation where low-mode averaging is applied to both
$lll$ and $hll$ contributions: in this case the large fluctuations are
sensibly suppressed. 
This confirms a posteriori that the fluctuations given by the $lll$ and
$hll$ contributions are the dominant ones.\\
At large time separations, we expect
\begin{equation}\label{corrfu}
C_N(t)=\bigg[A_{N}\left(e^{-m_{N} t}-e^{-m_{N}(T-t)}\right) +
 A_{N^*}\left(e^{-m_{N^*} t}-e^{-m_{N^*}(T-t)}\right) \bigg]...,
\end{equation}
where $m_N$ is the mass of the ground-state positive-parity nucleon, and
$m_{N^{*}}$ is the mass of the negative-parity state $N^{*}$ \footnote{Notice
that, because of our choice of the interpolating field, we cannot distinguish
between different parities, hence we have both contributions of $N$ and $N^*$
in both time sectors. Experimentally $\Delta_m=m_{N^{*}}-m_N\sim 600$ MeV, and we expect
that the contribution of $N$ in eq. (\protect\ref{corrfu}) dominates at
sufficiently large $t$.}.
In table \ref{masstab} we report the numerical results for the effective mass
at $t/a=9$, for our three values of the quark mass. 
For the local correlator, the fluctuations are such that a jackknife
statistical analysis of the errors cannot be trusted, and the uncertainties
reported in the first column for $am=0.04$ and $0.055$ are just for completeness.
For the lightest-quark mass, low-mode averaging applied on $lll$ and $hll$
with $n_b=20$ guarantees a reliable determination of statistical errors. A
brute comparison of the errors indicates a reduction by a factor 10, to be
compared with an increase of the computational cost by a factor 8. 
Going to heavier quark
masses the low-mode averaging becomes less efficient, as expected;
the same effect is foreseen by increasing the volume at fixed quark mass.
In fig. \ref{eff_mass} we report the results obtained for the
nucleon effective mass, for $am=0.04$. On the first figure on the left, no
low-mode averaging has been applied, and the signal for the nucleon is very poor. In the centre we
show the effect of applying low-mode averaging on the $lll$ part only: for
$t/a\gtrsim 7$ we already observe a significant improvement on the
signal. Finally, on the right we show the effective mass that we obtain after
applying low-mode averaging on $lll$ and $hll$ contributions; here the
statistical errors in the range $t/a\sim 9$--$12$ are under control and we can attempt
to extract the nucleon mass $m_N$.
It is important to stress that our uncertainty on the nucleon mass after low-mode
averaging is still quite large with respect to the errors obtained without
low-mode averaging but with different choices of the nucleon interpolating field
(see e.g. \cite{boston} for a recent computation).
For this reason, as a next step we intend to apply this method to baryon two-point functions with
other interpolating fields. Moreover, the technique can be easily generalized to baryon
three-point functions and could therefore be very useful for the computation of the electric
neutron dipole moment \cite{mass,Blum}.

\end{document}